# Enhancing Electron Coherence via Quantum Phonon Confinement in Atomically Thin $Nb_3SiTe_6$

J. Hu[1], X. Liu[1], C.L. Yue[1], J.Y. Liu[1], H.W. Zhu[1], J. B. He[2], J. Wei[1*], Z.Q. Mao[1*], L.Yu. Antipina[3,4], Z.I. Popov[5], P.B. Sorokin[3,4,6], T.J. Liu[7], P.W. Adams[7], S.M.A Radmanesh[8], L. Spinu[8], H. Ji[9] and D. Natelson[9],

[1]Department of physics and Engineering Physics, Tulane University, New Orleans, USA

[2]Coordinated Instrument Facility, Tulane University, New Orleans, USA

[3]Technological Institute for Superhard and Novel Carbon Materials, Moscow; Moscow Institute of Physics and Technology, Dolgoprudny, Russia Federation

[4]Emanuel Institute of Biochemical Physics, Moscow, Russian Federation,

[5]Siberian Federal University, Krasnoyarsk, Russia Federation

[6] National University of Science and Technology MISiS, Russian Federation

[7]Department of Physics and Astronomy, Louisiana State University, Baton Rouge, USA

[8]Advanced Materials Research Institute and Department of Physics, University of New Orleans, New Orleans, Louisiana 70148, USA

[9]Department of Physics and Astronomy, Rice University, Houston, USA

\* Address correspondence to jwei1@tulane.edu; zmao@tulane.edu

**The extraordinary properties of two dimensional (2D) materials, such as the extremely high carrier mobility [1,2] in graphene and the large direct band gaps in transition metal dichalcogenides $MX_2$ (M = Mo or W, X = S, Se) monolayers [3], highlight the crucial role quantum confinement can have in producing a wide spectrum of technologically important electronic properties. Currently one of the highest priorities in the field is to search for new 2D crystalline systems with structural and electronic properties that can be exploited for device development. In this letter, we report on the unusual quantum transport properties of the 2D ternary transition**

**metal chalcogenide- $Nb_3SiTe_6$. We show that the micaceous nature of $Nb_3SiTe_6$ allows it to be thinned down to one-unit-cell thick 2D crystals using microexfoliation technique. When the thickness of $Nb_3SiTe_6$ crystal is reduced below a few unit-cells thickness, we observed an unexpected, enhanced weak-antilocalization signature in magnetotransport. This finding provides solid evidence for the long-predicted suppression of electron-phonon interaction caused by the crossover of phonon spectrum from 3D to 2D [4].**

Recent advances in microexfoliation techniques [1] have made it possible to produce 2D atomic layered crystals that were previously inaccessible to the community. This has significantly expanded the breadth of research on low dimensional physics, as exemplified by the discoveries of exotic quantum phenomena in graphene [5] and $MX_2$ monolayers [6,7]. There is currently a significant effort in the field to extend the palate of 2D systems to the more complex ternary materials, with the expectation that the additional elemental complexity will extend the parameter space and possibly give rise to more exotic phenomena. In this paper, we report on the experimental observation of long-predicted suppression of electron-phonon interaction caused by 2D confinement in a new ternary 2D system - $Nb_3SiTe_6$ few-layer crystals.

$Nb_3SiTe_6$ was discovered two decades ago but scarcely studied [8]. As shown in Fig. 1a and 1b, the structure of $Nb_3SiTe_6$ is similar to that of $MX_2$ (*e.g.* $MoS_2$); both are formed from the stacks of sandwich layers, with comparable van der Waals (vdW) gaps. In $MX_2$, each X-M-X sandwich layer is composed of edge-sharing trigonal $MX_6$ prisms (Fig. 1c and 1d), whereas the Te-(Nb,Si)-Te sandwich layer of $Nb_3SiTe_6$ consists of face- and edge-sharing $NbTe_6$ prisms and Si ions insert into the interstitial sites among these prisms (Fig. 1a and 1b).

We have grown $Nb_3SiTe_6$ single crystals using chemical vapor transport; the lateral dimension of crystals can reach a few mm (see the inset in Fig. 1e). Sharp (0k0) x-ray diffraction peaks of these crystals (Fig.1e) confirm their layered structures and excellent crystallinity. Our resistivity measurements along in-plane ($\rho_{//}$) and out-of-plane ($\rho_{\perp}$) directions, as shown in Fig. 1f, indicate that $Nb_3SiTe_6$ is a quasi-2D metal, with the anisotropic ratio $\rho_{\perp}/\rho_{//}$ increasing from 9 at 300 K to 22.5 at 2 K (Fig. 1f, left inset).

Our DFT-PBE band structure calculations revealed that the metallicity of $Nb_3SiTe_6$ originates from the specific bonding state of Nb ions, as detailed in supplementary information. From the projected band structure and density of state shown in Fig. 2a, it can be seen that the valance bands crossing the Fermi level are derived from Nb-4d orbitals, indicating that the transport properties of $Nb_3SiTe_6$ are dominated by Nb-4$d$ electrons. Moreover, the decrease of dimensionality was found to lead to an unambiguous reconstruction of electronic band structure (Fig. 2b and 2c). When the thickness approaches single sandwich layer, the valance bands become much narrower and the gap between conduction and valence bands is doubled (~0.8 eV), while the Fermi level still crosses valence bands.

Atomically thin $Nb_3SiTe_6$ crystals can be obtained on $Si/SiO_2$ substrates via microexfoliation that has been widely used for graphene and $MX_2$ [1,9]. The flake thickness was first estimated by the color contrast under an optical microscope and the precise thickness was then measured by an atomic force microscope (AFM). Thin layers of ~3-5 nm are easily produced, and flakes as thin as only one unit cell (bi-layer) are accessible, as illustrated by the micrograph of a large (> 10 μm) bi-layer flake in Figs 3a and 3b.

We further characterized $Nb_3SiTe_6$ thin flakes through Raman spectra measurements and Transmission Electron Microscope (TEM) observations. As compared to $MX_2$ which displays only $E_{2g}^1$ and $A_{1g}$ phonon modes [10], $Nb_3SiTe_6$ exhibits more Raman modes (Fig. 3c) which may be attributed to its relatively complex lattice structure. Unfortunately, we are unable to identify these Raman modes due to the lack of theoretical studies of phonon spectra. However, we observed noticeable red- and blue- shifts caused by the decrease of flake thickness for 165 and 225 cm$^{-1}$ modes, respectively (see Fig. 3d). Often, such Raman mode shifts reflect the variation of phonon spectra with reducing dimensionality, as seen in MX$_2$ systems [10]. Our Raman measurements on < 4 nm flakes were unsuccessful since these flakes were easily damaged even by very low laser intensities. The stability of Nb$_3$SiTe$_6$ thin flakes were demonstrated by TEM observations. As shown in Fig. 3e, all the [010]-zone electron diffraction spots can be indexed according to the crystal structure of Nb$_3$SiTe$_6$, consistent with excellent crystallinity. Moreover, the atomic resolution image shows no visible amorphous structure (Fig. 3f) even after a few weeks exposure to ambient environment, confirming the stability of the few layers Nb$_3$SiTe$_6$.

To characterize the electronic properties of Nb$_3$SiTe$_6$ thin layers, we fabricated Nb$_3$SiTe$_6$ nano-devices using electron-beam lithography. Although we are able to thin Nb$_3$SiTe$_6$ crystals down to one-unit-cell, successful devices with good electrical contacts can be prepared only on flakes with the thickness ≥ 6 nm. The inset of Fig. 4b shows an image of a typical device. Overall, we observed a systematic increase of resistivity with the decrease of thickness, as shown in Fig. 4a. For relatively thick flakes (>12 nm), the resistivity presents a temperature dependence similar to that of the bulk sample. When the thickness is decreased below 12 nm, the resistivity displays marked changes in its temperature dependence. These thinner samples are more steeply temperature dependent and develop a low temperature upturn in the resistivity that is more pronounced with decreasing thickness.

In general, there are several mechanisms that can produce a low-temperature resistivity upturn, including Anderson localization, the Kondo effect and weak localization (WL). Through systematic measurements and analyses of the magnetoresistance (MR), we have determined that the observed resistivity upturn at low temperature is primarily attributable to Anderson localization, and not to the other two mechanisms. On one hand, the positive MR in all Nb$_3$SiTe$_6$ devices rules out the Kondo effect [11] and WL [12-15] (see Figs. 4b and 4c), phenomena associated with negative MR. On the other hand, with decreasing flake thickness, the field dependence of MR evolves from superlinear to sublinear behavior, with a zero field dip developing gradually (Fig. 4c). This reminisces the signature of weak antilocalization (WAL) in systems with strong spin-orbit coupling (SOC) [13-15], or disordered conductors with the electron-electron interaction (EEI) which causes corrections to density of states at the Fermi level [15,16]. Fortunately, these two mechanisms can be distinguished by the measurements of angular dependence of magnetoresistance (AMR). The EEI correction to conductance is not sensitive to the magnetic field orientation [16,17], while WAL is [15] (see the supplementary information). Our AMR data, together with the systematic analyses of MR data for all samples given below, point to a WAL scenario. As seen in Fig. 4d, when the applied field is low (*e.g.* 0.2T), the rotation of the field in the *x-z* plane (see the inset to Fig. 4d) leads to a striking dip in AMR at the direction where the field is parallel to the current. This anisotropic AMR reflects the suppression of WAL by the perpendicular field component $B_z$. Moreover, AMR becomes almost constant when the orientation angle of the field is beyond critical values ($\theta_c$), as indicated by the arrows in Fig. 4d. This implies that the Lorentz contributions to the MR are negligible at this field, and WAL is significantly suppressed for $B_z > B\cos\theta_c \sim 0.13$T. With increasing field, AMR gradually evolves to a $\cos^2\theta$ dependence, as shown by the fits of the data taken at 3T and 9T, indicating the orbital MR ($\propto B_z^2$) dominates.

While the observation of WAL in Nb$_3$SiTe$_6$ is not surprising due to relatively strong SOC induced by the heavy elements Nb and Te, the enhancement of WAL signature with reducing samples thickness is unexpected. WAL results from destructive quantum interference of electron waves, and causes enhanced conductivity [13-15], and therefore seems incompatible with Anderson localization. As we show below, their coexistence can be understood in terms of an anomalous enhancement of the WAL channel that is superimposed on the Anderson localization background.

To gain more insights into the thickness-dependent evolution of WAL in Nb$_3$SiTe$_6$ flakes, we have fitted the magnetotransport data using the Hikami-Larkin-Nagaoka (HLN) model [18], and extracted the quantum coherence length $l_\phi$ and spin relaxation length $l_{SO}$, which characterize the distance scales associated with quantum coherence and SOC, respectively (see supplementary information). With decreasing the sample thickness from 15 nm to 6 nm, the extracted $l_{SO}$ shows very weak thickness dependence. In contrast, $l_\phi$ increases by 50% (Fig. 4e) as thickness is reduced, implying that the observed enhancement of WAL is due to the larger coherence length in the thinner $Nb_3SiTe_6$ flakes.

We believe that the primary source of enhanced $l_\phi$ in our system is weakened inelastic electron-phonon (*e-ph*) scattering [13]. Other scattering channels such as electron-electron (*e-e*), electron-impurity, and interfacial scattering generally enhance with decreasing sample thickness [19,20]. These, of course, also contribute to the overall de-phasing rate, so we would naively expect $l_\phi$ to decrease or at least remain roughly constant as the sample thickness is reduced. Elastic scattering from static impurities, disorder, and interfaces can also result in a decrease of $l_\phi$ through their impact on the electronic diffusion constant. The fact that $l_\phi$ increases significantly as the system is brought through the crossover from 3D to 2D can only be attributed to a dimensional suppression of the inelastic *e-ph* scattering rate.

The detailed characteristics of *e-ph* interaction variation with thickness are further revealed by the temperature dependence of $l_\phi$, which is obtained from the fitting of magnetotransport data at different temperatures (see supplementary information). As shown in Fig. 4f, $l_\phi$ for 8, 10, and 15 nm samples can be well described by $l_\phi^{-2} = l_0^{-2} + AT^{P_e} + BT^{P_{ph}}$ [13], where $l_0$ represents a residual zero temperature coherence length, $AT^{P_e}$ and $BT^{P_{ph}}$ are associated with *e-e* and *e-ph* inelastic scatterings respectively. The fittings of $l_\phi(T)$ yield $P_e \sim 1$ for all samples (see Tab. 1), consistent with the prediction of the localization theory [13]; The value of $P_{ph}$ sharply decrease from 3.68 for the 15nm device to 2.21 for the 8 nm one, suggesting the variation of *e-ph* interactions with the flake thickness. More importantly, the weight of *e-ph* interaction (*i.e.* the prefactor B in Tab. 1) is reduced by two orders of magnitude in the 8 nm sample as compared to the 15 nm flake.

In lower dimensions, the *e-ph* interaction can be modified by the quantum confinement of both the electronic band structure and the phonon spectrum. As shown in Fig. 2, the reduction of thickness leads to significant electronic band flattening in monolayer. Band narrowing is predicted to suppress *e-ph* interaction [21,22]. However, our band structure calculations did not show significant band narrowing in bi-layer and thicker $Nb_3SiTe_6$ flakes (see Fig. 2b). Therefore, the suppression of *e-ph* scattering in our thinner samples cannot be attributed to 2D confinement on electronic band structure. Instead, we believe that confinement plays a much more important role in its modification of the phonon spectrum, which in turn affects the coherence length. As we show in the supplementary information, the phonon wavelength estimated from both the experiments and *ab initio* calculations is ~ 80-100 nm at 2K, much larger than the thickness of those devices showing enhanced WAL, suggesting the phonon spectra crossover from 3D to 2D.

It has already been proposed that the 2D quantum confinement on phonons may lead to the opening of a gap for acoustic phonons [4,23] and therefore suppress *e-ph* interactions [4]. Traditional experimental probes of phonon properties, such as heat capacity and thermal conductivity, are difficult to perform on mesoscopic samples. Furthermore, electron heating measurements on a variety of 2D systems have produced contradictory results [13,24,25], and the role of dimensionality on *e-ph* scattering remains unclear. On the other hand, quantum interference are extremely sensitive to inelastic scattering [13]. Earlier quantum transport studies on ultra-thin films did not find the evidence for the suppression of *e-ph* scattering with reducing dimensionality [26-29]. However, those studies were made on polycrystalline films that typically have a significant grain-boundary scattering rate, distinct from our crystallized atomically thin $Nb_3SiTe_6$ layers. In addition, those materials may be much more sensitive to interfacial scattering due to their isotropic lattice structures. In contrast, the highly anisotropic sandwich-like lattice structure of $Nb_3SiTe_6$ leads to weak interlayer electronic coupling (Fig. 1f). Thus the scattering from the crystal/substrate interface, which is likely responsible for our observed resistivity upturn at low temperature, may be quickly "screened" by the adjacent few layers near the substrate so that WAL can occur in layers further away from the interface.

To summarize, we have produced atomically thin analogs of the ternary chalcogenide $Nb_3SiTe_6$. Using quantum interference as a probe of the electron-phonon scattering rate, we find clear evidence for the long-predicted dimensionality suppression of *e-ph* interactions that resulted from the crossover from a 3D to 2D phonon spectrum. Such phonon dimensionality effect may also play an important role in the quantum transport properties of other layered crystalline materials.

**Methods**

The $Nb_3SiTe_6$ bulk single crystals were synthesized with a stoichiometric mixture of starting elements using the chemical vapor transport. During the growth the temperature was set at 950 ˚C and 850 ˚C, respectively, for hot and cold ends of the zoned tube furnace. The composition and structure of these single crystals was confirmed using Energy-dispersive X-ray spectroscopy and X-ray diffraction measurements. To avoid any possible secondary phase in our single crystals, in addition to the careful characterization of bulk single crystals, we double-checked the tiny pieces of single crystals on the scotch tapes used during the exfoliation. For TEM measurements, $Nb_3SiTe_6$ thin flake was transferred to a Ni/C grid. The $Nb_3SiTe_6$ nano devices were fabricated using standard electron-beam lithography followed by a deposition of Ti(5nm)/Au (50nm). The transport measurement for both bulk and nano devices are performed in a physical properties measurement system (PPMS).

The electronic band structure calculation was performed using density functional theory in the framework of generalized gradient approximation (GGA) in Perdew-Burke-Ernzerhof [30] parameterization with periodic boundary conditions using Vienna Ab-initio Simulation Package [31]. Projector-augmented wave method along with a plane wave basis set with energy cutoff of 220 eV was used. To calculate equilibrium atomic structures, the Brillouin zone was sampled according to the Monkhorst–Pack [32] scheme with a *k*-points mesh 3×6×4. To avoid spurious interactions between neighboring structures in a tetragonal supercell, a minimum of vacuum layer of 20 Å in all non-periodic direction was included. Structural relaxation was performed until the forces acting on each atom were less than 0.05 eV/Å.

**Acknowledgements**

The authors are grateful to John DiTusa for informative discussions. The work at Tulane is supported by the NSF under grant DMR-1205469 and the LA-SiGMA program under award #EPS-1003897. PWA and TJL acknowledge the support of the U.S. Department of Energy, Office of Science, Basic Energy Sciences, under Award No.DE-FG02-07ER46420. LYA and PBS acknowledge the support of the Russian Science Foundation (project #14-12-01217) and grateful to Joint Supercomputer Center of the Russian Academy of Sciences and "Lomonosov" research computing center for the possibilities of using a cluster computer for the quantum-chemical calculations.  DN and HJ acknowledge support through U.S. Department of Energy, Office of Science, Basic Energy Sciences award DE-FG02-06ER46337. The work at UNO is supported by the National Science Foundation under the NSF EPSCoR Cooperative Agreement No. EPS-1003897 with additional support from the Louisiana Board of Regents.

**Author contributions**

J.H, J.Y.L., H.W.Z., and Z.Q.M. carried out bulk sample growth and characterization including XRD, resistivity and specific heat measurements. J.H, X.L., C.L.Y., and J.W. fabricated nano devices. J.H, X.L., T.J.L., P.W.A., S.R, and L.S. collected resistivity and magnetotransport data for nano devices. J.H and J.B.H. carried out TEM measurements. H.J and D.N. performed Raman spectrum measurements. L.Y.A, Z.I.P, and P.B.S. calculated the electronic structure. J. H, J.W., Z.Q.M., P.W.A, D.N and P.B.S analyzed the data and wrote the manuscript. J.H and X.L. contributes equally to this work.

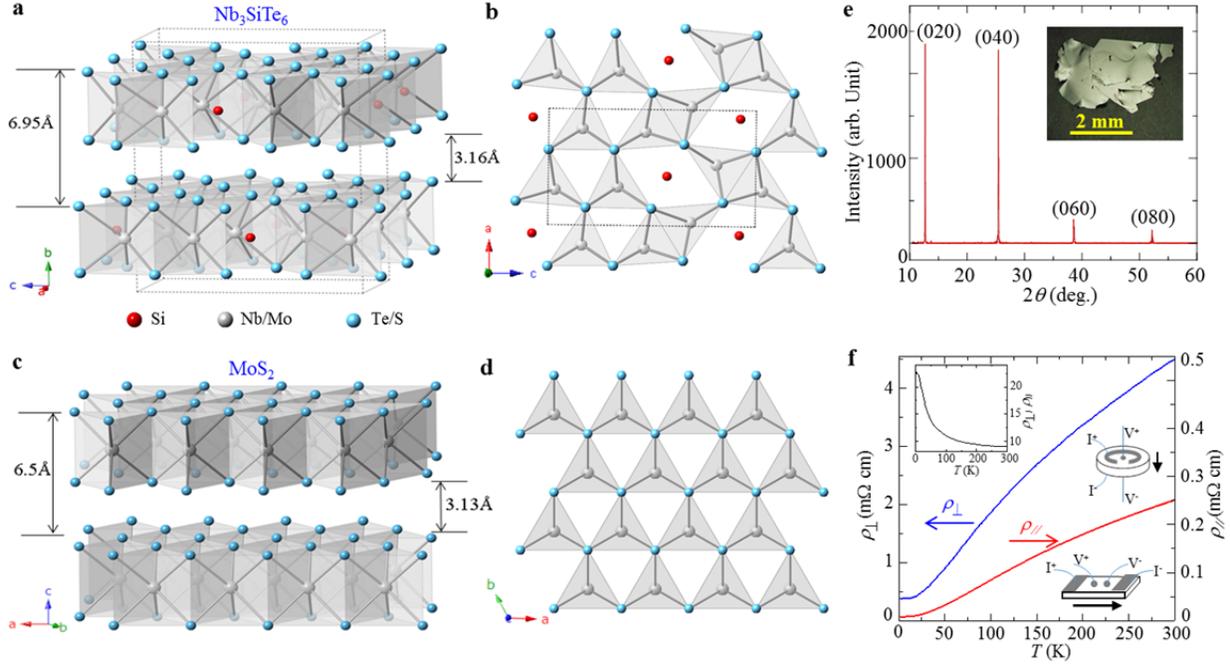

**Figure 1 | Structure and electronic properties of bulk Nb$_3$SiTe$_6$. a** and **c**, The crystal structures of (**a**) Nb$_3$SiTe$_6$ and (**c**) MoS$_2$. Note that the out-of-plane axis is defined to be the *b*-axis for Nb$_3$SiTe$_6$. **b** and **d**, The top view of a single layer of (**c**) Nb$_3$SiTe$_6$ and (**d**) MoS$_2$. **e**, Single crystal X-ray diffraction pattern along the (0k0) plane for bulk Nb$_3$SiTe$_6$. The out-of-plane lattice constant *b* is determined to be ~1.4 nm, consistent with the previously-reported structure with *b* = 1.3938(5) nm [8]. Inset in panel **e**: an optical image of a bulk Nb$_3$SiTe$_6$ single crystal. **f**, Temperature dependences of the in-plane ($\rho_{//}$) and out of plane ($\rho_\perp$) resistivity for the bulk Nb$_3$SiTe$_6$ single crystals. Left inset: the anisotropic ratio ($\rho_\perp/\rho_{//}$) as a function of temperature. Right insets: The contact lead configurations for $\rho_\perp$ (upper) and $\rho_{//}$ (lower) measurements.

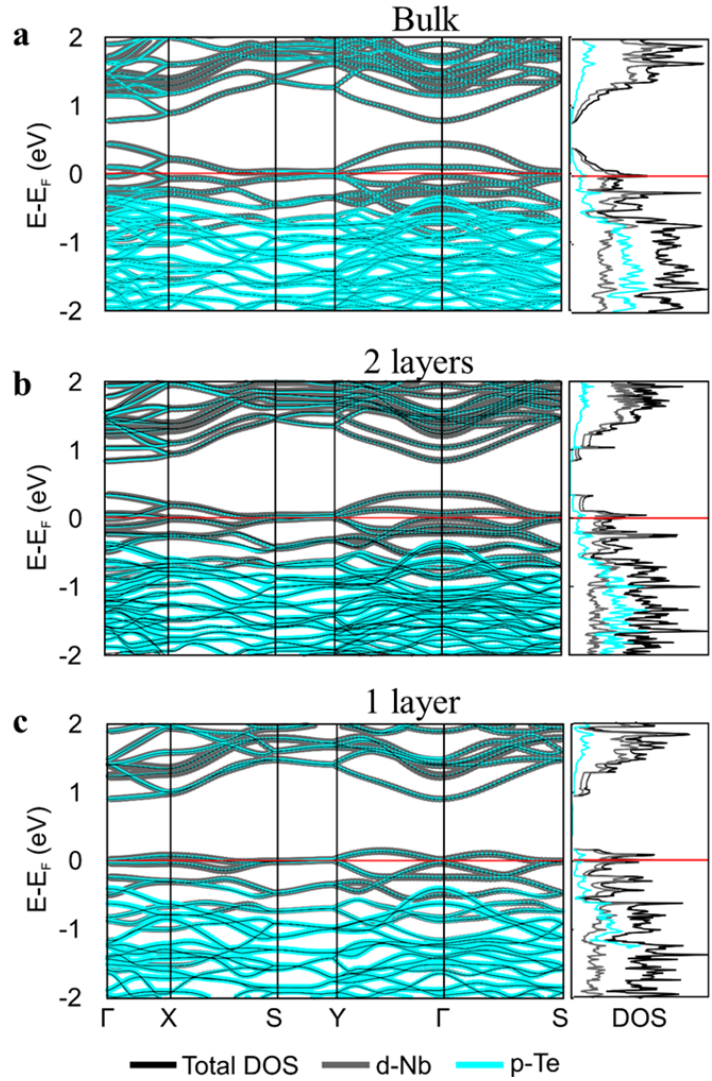

**Figure 2 | a-c**, Projected electronic band structure and density of states of $Nb_3SiTe_6$ for (**a**) bulk, (**b**) bi-layer, and (**c**) mono-layer. The gray and cyan curves in the left panels represent energy bands derived from Nb-4$d$ and Te-5$p$ orbitals respectively, while the black curves in the right panels represent total density of states. The Fermi energy is marked by the horizontal red lines.

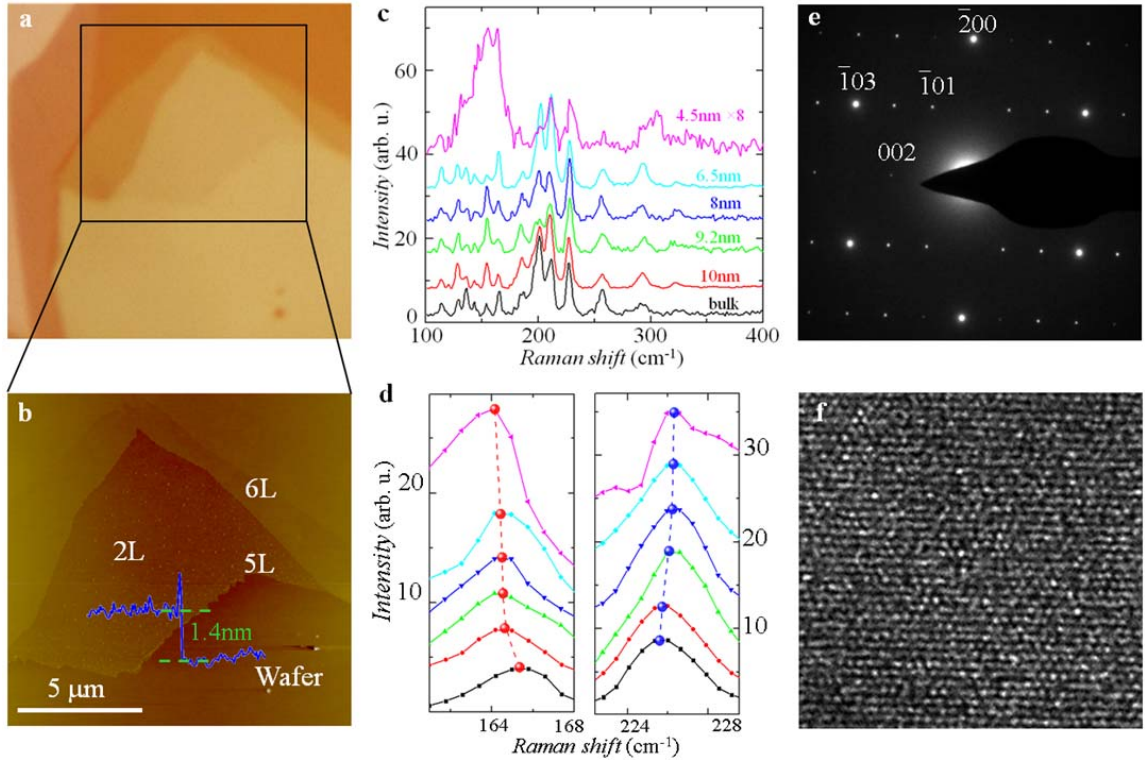

**Figure 3 | AFM, TEM and Raman observation of $Nb_3SiTe_6$ atomically thin layers. a**, Optical image of a $Nb_3SiTe_6$ thin flake. **b**, AFM image of a selected area from Image **a;** 2L, 5L and 6L represent bi-layers, 5-layers and 6-layers respectively. The inset shows the height of the bi-layer (~1.4nm) from the profile scan. **c**, Raman spectra of $Nb_3SiTe_6$ crystals with various thickness (displaced vertically). The spectrum for the 4.5 nm flake is multiplied by a factor of 8, in which the hump around 150 nm$^{-1}$ is caused by laser damage of the $Nb_3SiTe_6$ crystals. **d**. Enlarged spectra for Raman peaks around 165 and 225 cm$^{-1}$ for various thickness (displaced vertically). **e**, [010] zone (perpendicular to layers) selected area electron diffraction (SAED) pattern of a $Nb_3SiTe_6$ thin flake. The overall rectangle arrangement of diffraction spots is consistent with the orthorhombic structure of $Nb_3SiTe_6$, while the hexagonal pattern of brighter spots reflects the hexagonal Te lattice plane. **f**, High resolution transmission electron microscopy image of the thin flake shown in **c**.

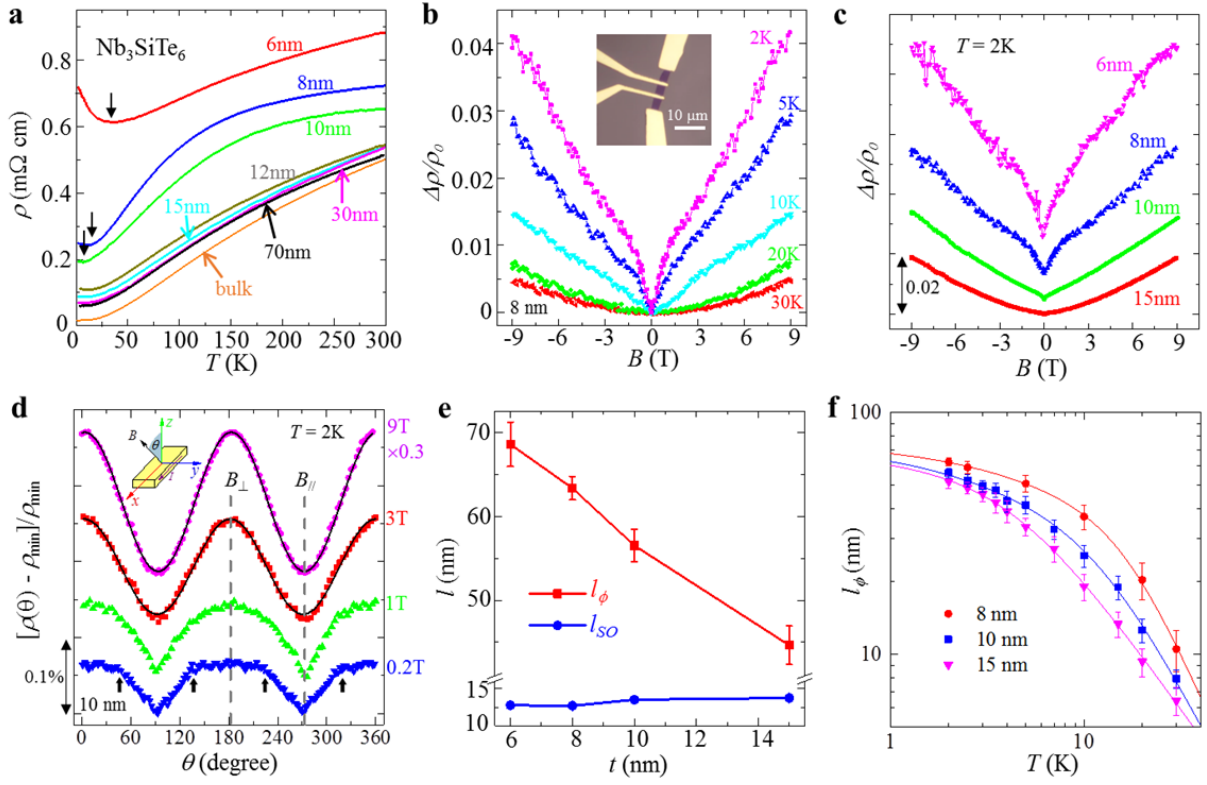

**Figure 4 | Transport properties of Nb$_3$SiTe$_6$ nano-devices**. **a**, Temperature dependence of resistivity of Nb$_3$SiTe$_6$ thin flakes with various thicknesses. The bulk resistivity data is adopted from Fig. 1f for comparison. The downward arrows mark the resistivity upturns. **b**, Normalized magnetoresistivity, $\Delta\rho/\rho_0 = [\rho(B)-\rho(B=0)]/\rho(B=0)$, of a 8 nm Nb$_3$SiTe$_6$ flake. The inset shows the optical image of this device. The magnetic field was applied along the out-of-plane direction. **c**, Normalized magnetoresistivity $\Delta\rho/\rho_0$ at $T$ = 2K for Nb$_3$SiTe$_6$ flakes of various thicknesses. Data for different thickness are displaced vertically for clarity. **d**, Isothermal angular magnetoresistance $\Delta\rho/\rho_{min} = [\rho(\theta)-\rho_{min}]/\rho_{min}$ at $T$ = 2K under various magnetic fields for the 10 nm device. $\rho_{min}$ is the resistance minimum which occurs for ***B***//***I***. The angle $\theta$ is the angle between the out-of-plane axis ($z$-axis) and the magnetic field vector. The AMR at 9T is multiplied by a factor of 0.3. The solid line shows the fits of the 9T and 3T data to the $\cos^2\theta$ dependence. Inset: schematic of the measurement setup. The magnetic field rotates within the x-z plane defined by the current ($x$-axis) and out-of-plane axis ($z$-axis). **e,** The coherence length $l_\phi$ and spin-orbit relaxation length $l_{SO}$ extracted from fitting (see supplementary information for details). The error bars for $l_{SO}$ is too tiny to

be seen. **f**, Temperature dependence of coherence length $l_\phi$ for the 8, 10, and 15 nm samples. The solid lines show the fits for $l_\phi$ (see text).

**Table 1** Parameters of *e-e* and *e-ph* interactions obtained from the fits of the temperature dependence of $l_\phi$.

|  | $l_0$ (nm) | A | B | $P_e$ | $P_{ph}$ |
|---|---|---|---|---|---|
| 8nm | 74.54 ±1.12 | 4.65(±0.10)×10$^{-5}$ | 2.47(±0.37)×10$^{-8}$ | 1.02(±0.05) | 3.68(±0.21) |
| 10nm | 73.74 ±1.75 | 6.21(±0.31)×10$^{-5}$ | 3.86(±0.27)×10$^{-7}$ | 1.01(±0.02) | 3.11(±0.13) |
| 15nm | 72.07 ±1.27 | 6.03(±0.30)×10$^{-5}$ | 9.23(±0.35)×10$^{-6}$ | 0.95(±0.07) | 2.21(±0.11) |